\documentclass[11pt]{article}
\usepackage{moriond,epsfig}

\def\be{\begin{equation}}
\def\ee{\end{equation}}
\def\bea{\begin{eqnarray}}
\def\eea{\end{eqnarray}}

\def\cygx{Cyg~X$-$1}

\begin{document}
\vspace*{4cm}
\title{HIGH ENERGY EMISSION FROM GALACTIC \\
BLACK HOLE SYSTEMS}

\author{A. GOLDWURM}

\address{Service d'Astrophysique, DAPNIA/DSM/CEA-Saclay, 91191 Gif sur Yvette, France}

\maketitle\abstracts{
In the last decade our knowledge on galactic black hole systems
and in particular on their high energy behavior has considerably improved.
I will briefly review here the main results obtained 
by the high-energy missions SIGMA/GRANAT, Compton-GRO, Beppo-SAX and Rossi-XTE,
on these objects and, in particular,
I will discuss the spectral shapes observed at energies higher than 30~keV 
and the detections of high energy features at $>$ 300~keV.
Galactic black holes are indeed main targets for the ESA gamma-ray
mission, INTEGRAL, to be launched in October 2002, 
and for the next gamma-ray missions SWIFT, AGILE and GLAST.
We expect that a large amount of data will be collected in the next 
years and the perspectives for the high energy astrophysics of 
galactic black hole systems look very promising.
}

\section{Black Holes in Accreting Binary Systems}

Our Galaxy probably hosts a large number of black holes (BH) of stellar mass 
sizes.
Black holes, by definition, are difficult to observe, unless they are associated
to a normal star in a close binary system. When in such a binary system 
transfer of matter occurs from the companion star to the black hole, 
the system may become a powerful source of X-rays.
In the last 12 years (1990-2002) our knowledge of galactic 
black hole binary systems 
has increased enormously thanks to the large amount of results 
obtained in the standard X-ray band with 
Ginga, Rosat, ASCA, and now Chandra and XMM-Newton,
and in the hard X-ray and gamma-ray range
with SIGMA/GRANAT, Compton-GRO, Beppo-SAX and Rossi-XTE.
Observations have shown that X-ray sources which seem
associated to black holes in binary systems are very hard sources
and infact often appear the brightest objects at energies between 
10 keV - 1 MeV.
They are also often observed to emit radiowaves 
and some of them, the so called {\it microquasars}, have been associated 
to radio-jet sources (Mirabel $\&$ Rodriguez 1999).

These systems are powered by accretion of matter provided by the 
companion star, which releases, by falling into the deep potential 
well of the hole, gravitational energy in form of high energy 
radiation.
These systems are classified as high mass X-ray binaries (HMXB), 
when the star is young, massive ($>$~1~M$_{\odot}$), 
of early spectral type (O, B), 
or as low mass binaries (LMXB), when the secondary is old, of low mass 
($<$~1~M$_{\odot}$), and of late spectral type (later than A). 
HMXB are generally large systems with orbital periods P$_o$~$>$~5 days 
and the accretion usually occurs by the capture of the material from
the strong stellar wind of the companion.
It is not always clear whether a disk forms.
In the small, low period LMXB (P$_o$ $\approx$ few~hr - 10~days), instead,
the secondary fills the Roche lobe and matter is accreted with large angular 
momentum, often through an accretion disk.

All these objects are variable but some are visible most of 
the time ({\it persistent sources}) while others rather spend most of their
life in a quiescence state, where they are very faint and even undetectable
({\it transient sources}). 
Obviously the classification depends slightly on
the level of instrument sensitivity and
the boundary between persistent and transient systems is somehow 
arbitrary. Some transients may reach a state where they are active
for long time displaying erratic flux variability 
(e.g. GRS~1915+105) and some persistent sources may pass large period 
in very low states (e.g. GRS~1758-258, GX 339-4).

Transients are particularly interesting because during their quiescence 
period, their mass function can be measured in detail.
Optical and infrared (IR) spectro-fotometry observations of binary systems 
in suitable conditions 
can provide radial velocity measures of the secondary orbital motion,
which allows determination of the mass function of the system. 
We presently know 14 binary systems in our Galaxy and in the 
Magellanic Clouds, whose compact object mass has been determined 
(or constrained) and results larger than 3~M$_{\odot}$,
the theoretical mass limit of a stable neutron star (NS), 
implying the presence of a black hole.
Out of these dynamically proven BH, 3 are HMXB, 
the persistent sources \cygx, LMC~X-1, LMC~X-3, the rest are
all X-ray Novae.
X-Ray Novae (XN), also known as Soft X-ray Transients,
are LMXB normally in quiescence state, 
which undergo sudden few-month-long X-ray outbursts with 
typical recurrent times of few tens of years (Tanaka $\&$ Shibazaki 96). 
After the flare the systems generally return to the quiescence phase,
the disk contribution to optical and IR emisson is reduced and
the companion (identified during the outburst) can be studied in detail.
That is why many XN have known mass functions and most of them are found 
to harbour a black hole as primary compact object.
The most recent determination of a GBH mass is the one for GRS~1915+105
the first discovered superluminal microquasar, which contains
the most massive BH ($\approx$ 14~M$_{\odot}$) of the known GBH 
(Greiner et al. 2001).

When the mass function cannot be determined, X-ray sources can be identified 
as BH candidate systems on the basis of their X-ray and gamma-ray
spectral and variability properties. In particular the lack of distinctive 
signs of NS system variability (coherent pulsation, 
type I and II X-ray bursts) and the presence of a strong hard spectral component 
extending in the range $>$~30~keV (plus signs of fast variability 
and variable ultrasoft component) are used to search for BH candidates.
Several X-ray sources identified as BH candidates using a spectral criterium 
were later confirmed as BH systems from mass function measures.
About 15 sources are considered today on these basis very serious BH 
candidates. Three of them are persistent sources, 
the variable source GX 339-4 and the two microquasars of the Galactic Center 
region 1E~1740.7-2942 and GRS~1758-258, probably LMXB for which no optical 
counterpart has been found. The others are member of the X-ray novae class.

\section{High Energy Spectra of GBH}

The prototype of the GBH class is the persistent, very bright and well known
source \cygx. 
It is a HMXB with a P$_o$ of 5.6 days, a O9.7 Iab companion and 
an estimated BH mass of more than 6~M$_{\odot}$.
Accretion takes place via the focused stellar wind of the supergiant,
generating intense and variable X-ray emission. 
After its discovery it was remarked that 
its peculiar spectrum extending in the hard X-ray range with a rather 
flat power law slope (photon index $\alpha \approx$ 1.5-2.0)
and an exponential decay at energies $>$ 50-100 keV,
was remarkably different from the spectra of neutron star binary systems,
and could not be explained by thermal models, in particular
by those describing optically thick disks.
This emission is generally interpreted as due to repeated inverse Compton 
scattering (Comptonization) of soft photons by the thermal energetic electrons 
of a hot (kT $\approx$ 50-100 keV) cloud (or corona) with 
typical scattering optical depth $\tau~\approx$~1. 
Sunyaev $\&$ Titarchuk (1980) described the emergent spectrum 
in the approximation of $\tau >$~1 and kT$<<$~mc$^2$, while
later Titarchuck (1994) provided analytical expressions for relativistic case
and larger range of kT and $\tau$.
Observations by ART-P and SIGMA/GRANAT of \cygx\ spectrum in the range 3-300 keV 
indeed showed the limitation of the Sunyaev $\&$ Titarchuk (1980) 
approximation (Grebenev et al. 1993) but the data could still be fitted 
with more realistic approximations of the Comptonization model.

\cygx\ also occasionally shows a different spectrum which is dominated by a 
strong ultrasoft thermal component which peaks at energies $<$~1-2~keV 
and which is generally interpreted as the multicolor black body 
emission from a geometrically
thin and optically thick disk as described by Shakura $\&$ Sunyaev (1973). 
In this spectral state, the power law component is much weaker 
and steeper ($\alpha \approx$ 2.5-3.0) and in some cases absent. 
In general the variable soft component is found to be
anticorrelated to the hard component.

The standard interpretation of these spectra is that the seed photons of
the Comptonization come from the disk and, when the accretion rate 
increases, the emission from the disk also increases, generating 
the soft thermal component. 
The large soft photon flux cools down by Compton scattering the hot cloud 
and quenches the hard component.
In the last ten years these two different spectra
have been observed in many other BH systems 
(e.g. Grebenev et al. 1993, Grove et al. 1998, Fig. 1) 
and infact detection of these two spectral components 
can be used to identify accreting binaries which probably contain a BH 
rather than a NS.
In spite of the fact that NS binaries show occasionally hard tails
or soft components, in general we now know how to recognize BH from 
from their high energy spectra.
The large amount of data collected from these sources have also 
allowed to characterize their spectral/variability states 
and the open questions now concern 
the physical origin and transitions of these states.

\begin{figure}
\centering{
\epsfig{figure=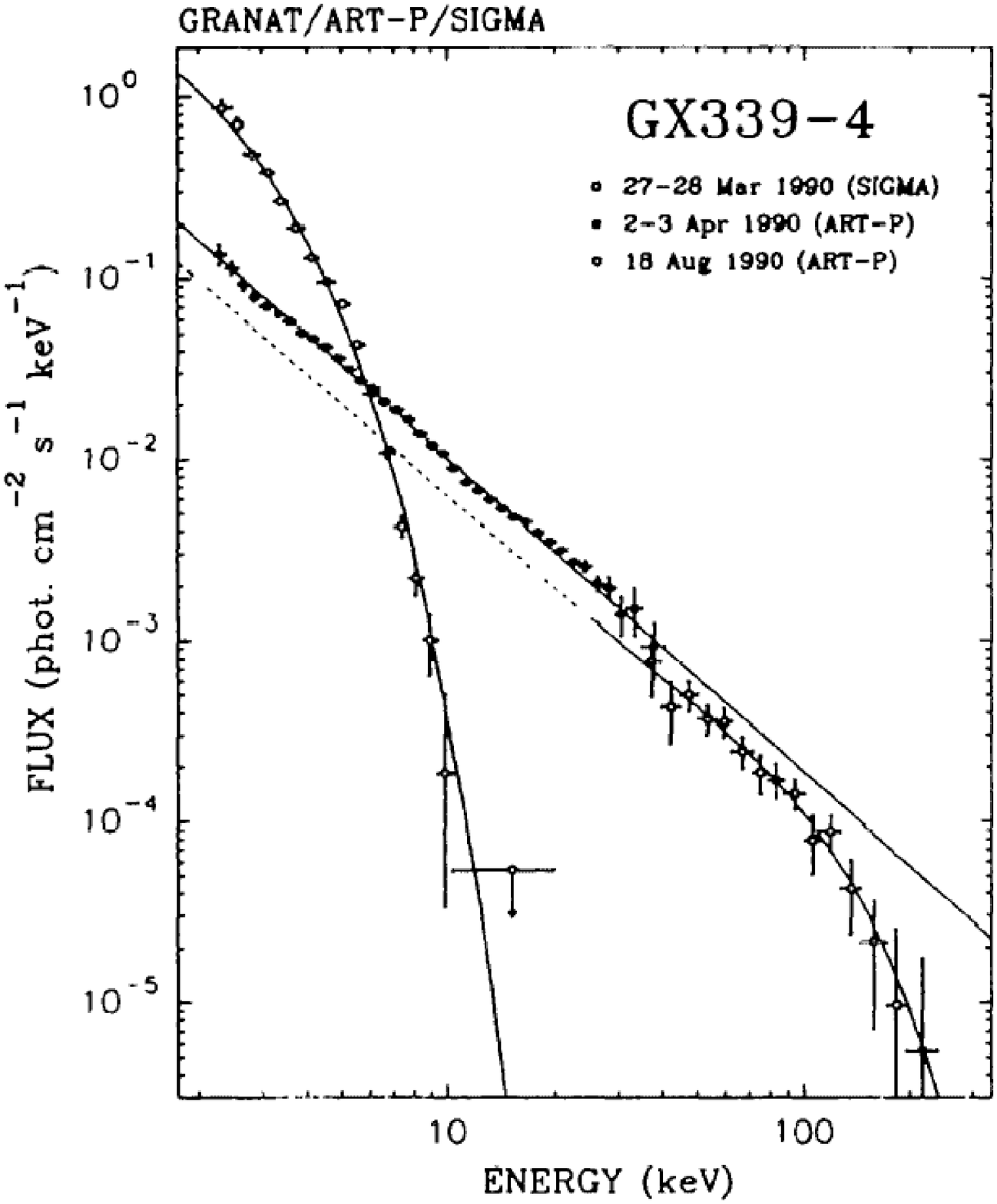, width= 6 cm} ~~~~~~~~~~~~~
\epsfig{figure=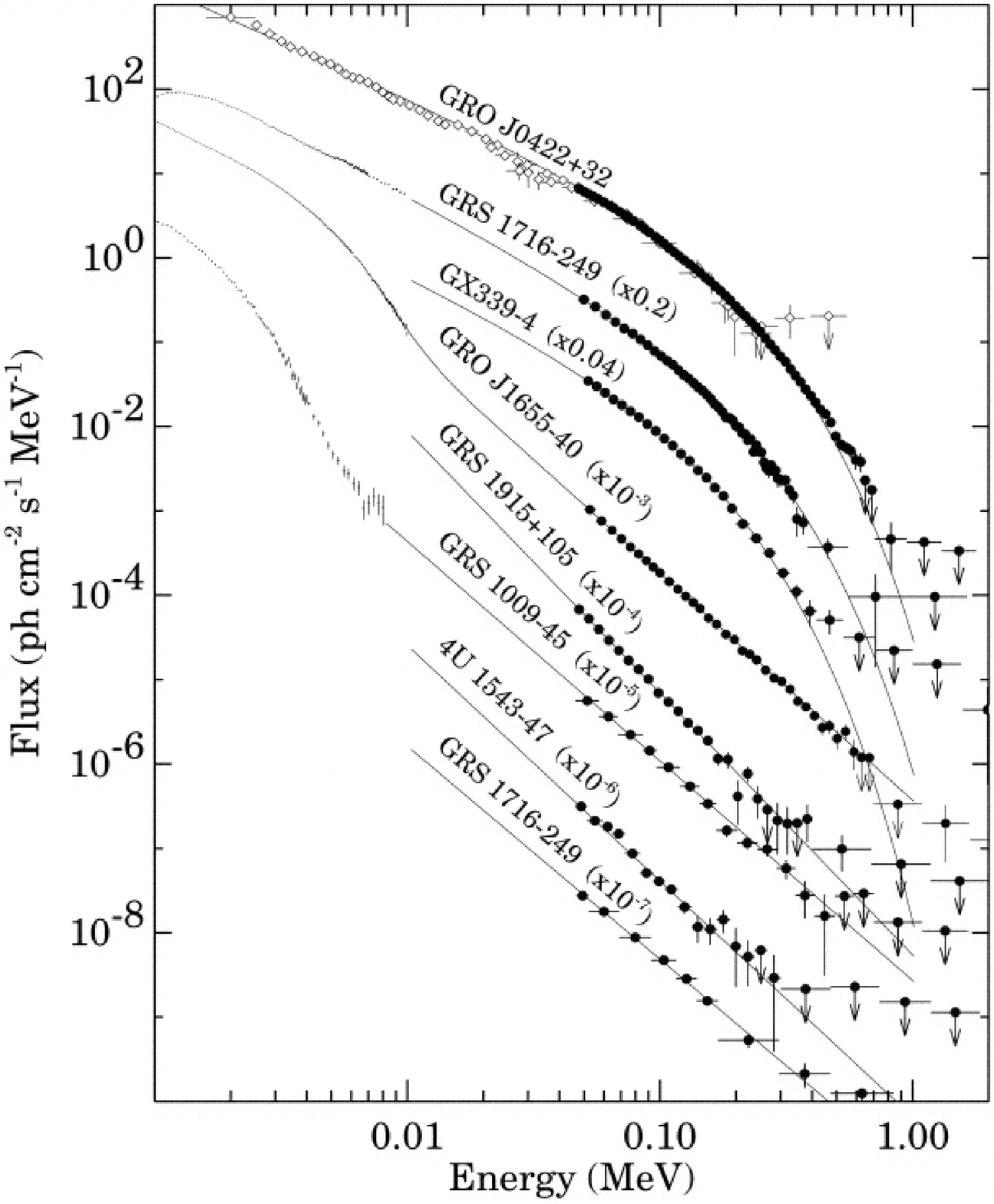, width= 6 cm} 
}
\caption{GRANAT/SIGMA-ARTP spectrum of GX~339-4 (from Grebenev et al. 1993) 
(left). Combined ASCA, TTM, HEXE and Compton-GRO spectra of GBH 
(from Grove et al. 1998) (left).
\label{fig:gbhspectra}}
\end{figure}

\section{Black Hole Spectral States, State Transitions and Fast Variability}

{\bf Five canonical states} have been identified in BH binaries:
the Very High State (VHS), High/Soft State (HS), Intermediate State (IS), 
Low/Hard State (LS) and Quiescence State (van der Klis 1995, 
Mendez et al. 1999). 
These states are characterized by different combination of the ultra-soft, 
hard and reflection (continuum excess, fluorescence lines and K-edge) 
components (Fig. 2 left) coupled to different properties of variability 
and QPOs (Fig. 2 right).
The HS is characterized by the strong ultra-soft thermal
component along with a very weak and steep (or even absent) power-law extending 
to high energies.
The spectrum of the LS is instead dominated by the hard power law 
with exponential cut-off as described above.
The VHS resembles the HS where the hard component, though steep, is 
bright (Fig. 1 right) and it extends to high energies 
without evidence of clear exponential break up to $>$ 200-300 keV.
The IS is similar to the VHS (both soft and hard component are present) 
but with a much lower level of emission.
Quiescent state is observed in XNs and it is characterized by a very low 
luminosity ($<$~10$^{33}$~erg~s$^{-1}$).

\begin{figure}
\centering{
\epsfig{figure=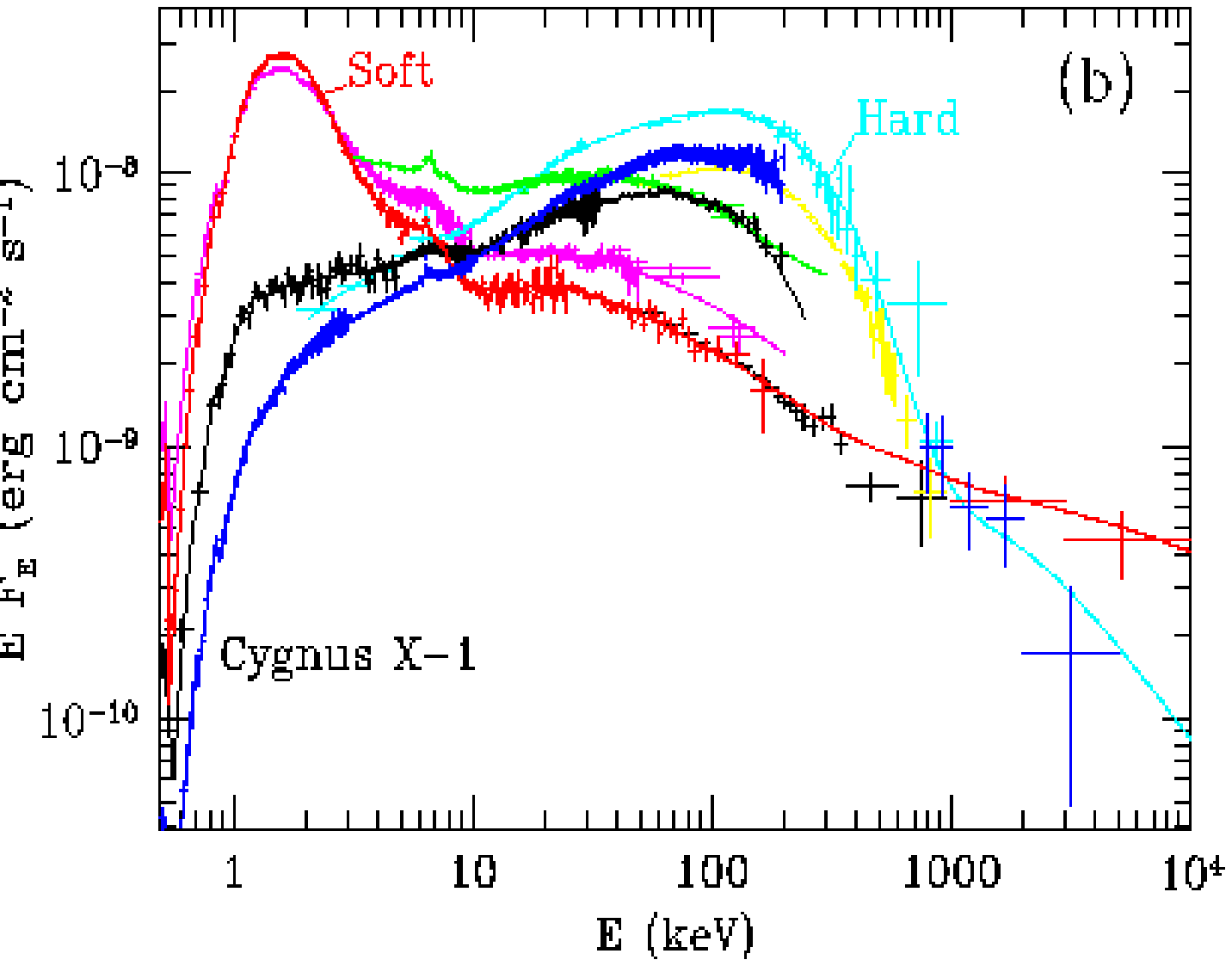, width= 7.5 cm} ~~~~~~~~~
\epsfig{figure=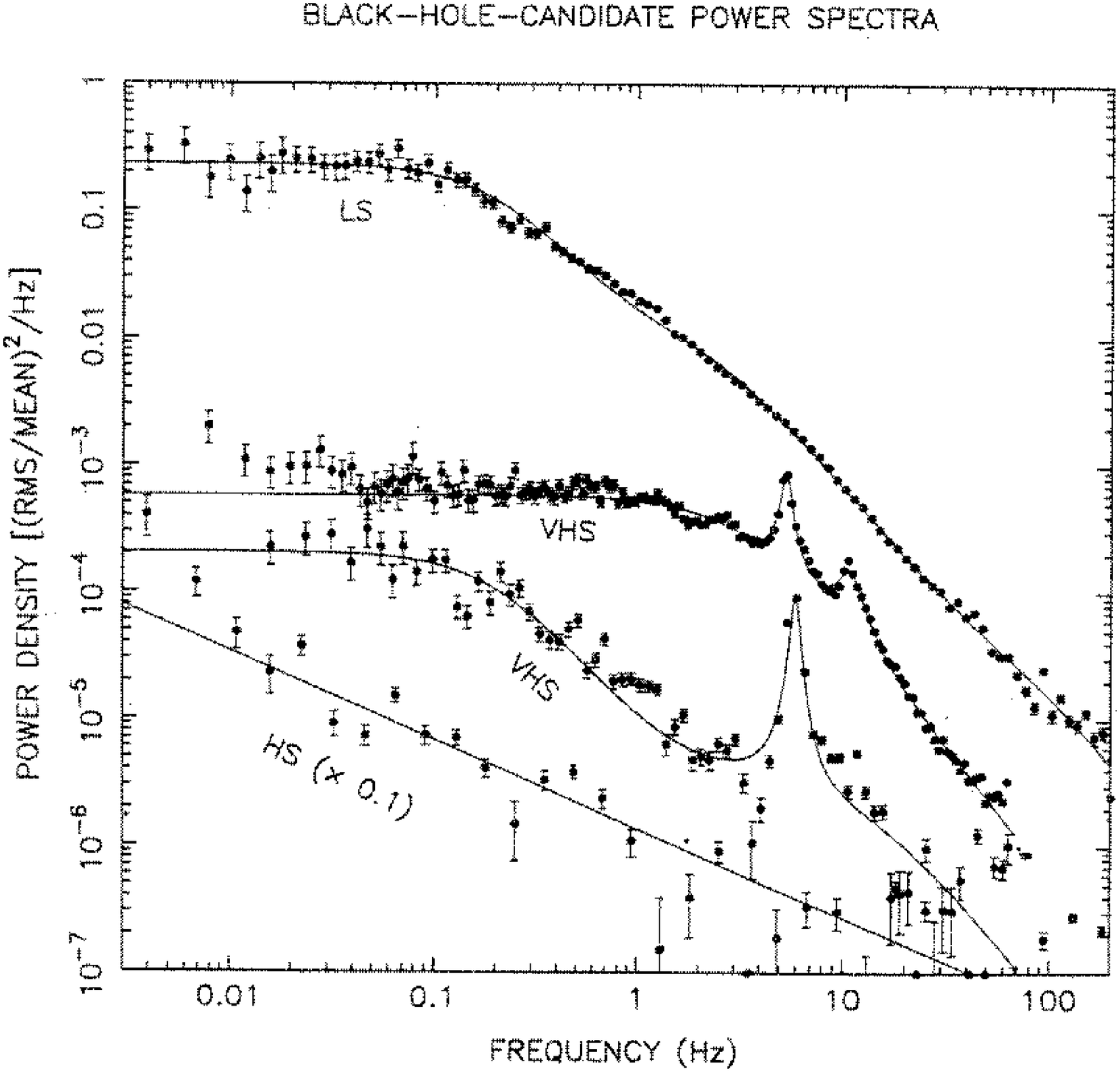, width= 6.5 cm}
}
\caption{Broad-band spectra of \cygx\ in different spectral states
(from Gierlinski et al. 1997, 1999) (left).
Power spectral density of GBH candidates in different spectral states.
From top to bottom: \cygx\ in HS, Musca XN in VHS, IS and HS.
(from van der Klis 1995) (right).
\label{fig:gbhvar}}
\end{figure}

During this phase, material is transferred 
from the companion at a very low rate and it is stored 
in a low-viscosity accretion disk, 
without effective accretion into the compact object. 
The XN outburst is probably initiated by a thermal instability 
which suddenly produces the fall of the matter accumulated into the
disk during quiescence.
The effective accretion rate rises to a maximum in few days 
and decreases over several weeks (often with exponential law)
spanning several orders of magnitudes.
Thus XN experience different accretion rates and they are often seen 
over a period of a month or two, to display the different GBH 
states described above.
At the onset of a XN outburst the hard X-rays usually reach 
the maximum first and then the soft-component develops, the XN enters the
very high state, the high/soft state 
and tends to evolve into the intermediate and the low/hard states 
before entering again the quiescence phase. 
The sequence of states seems driven by the accretion rate variation.
However, this pattern is not always followed, for example
some, even very bright, X-ray Novae have been observed throughout 
the whole outburst in low/hard state only 
(e.g. GRO J0422+32 and GRS 1716-249, Fig. 1). 

In addition to long term variations, GBH also show rapid aperiodic variability
on timescales from several hundred seconds down to milliseconds.
BH spectral states are also associated to the specific properties of
the fast variability, which is generally described by the power spectral
density (PSD) (Fig. 2 right) (van der Klis 1995).
In the HS the GBH PSD show a large level of variability (20-35~$\%$ rms),
with flat slope (red noise) at low frequencies up to a break frequency $\nu_B$ 
($\approx$~0.1-1~Hz) and with a power-law slope of index $\approx~-$1 (flickering)
at higher $\nu$.
In the HS the level of variability is very low ($<$~5$\%$) and the PSD has a 
steep power-law slope. In VHS (and IS) two kind of power spectra are observed 
one similar to the LS but with lower level of integrated rms ($<$ 15-20~$\%$)
and with $\nu_B~>$~1~Hz, the other spectrum rather similar to the HS PSD.
Features in the PSD, or quasi periodic oscillations (QPO), 
have also been detected in GBH. 
Broad QPOs, sometimes referred as peaked noise, are seen at low frequency 
($<$ 1 Hz) in LHs, while strong and narrow QPO are observed between 
1-10 Hz in VHS.
Characteristic times ($\nu_B$, $\nu_{QPO}$) and level of noise seem to vary 
with source intensity, photon energies and in correlation between them.
One important feature which is generally observed is that
hard photons lag soft photons with typical time lags of 
milliseconds-seconds. 
While the Comptonization can explain in principle the lags, 
the observed dependency of the lags with frequency requires non homogeneous
Comptonizing clouds.
The physical link between spectral state and observed variability in BH
binaries in still not understood, in spite of the remarkable 
progresses obtained in the recent past thanks in particular to RXTE results,
and the relative succes of shot noise models to interpret the observed PSD
(Poutanen et al. 2001).

\section{Models and High Energy Tails}

Different models of accretion flows have been proposed to account for 
the BH spectra and their state transitions (see e.g. Liang 1998). 
They differ mainly on the
origin and geometry of the comptonizing plasma producing the hard component.
One popular scenario is that 
the optically thick accretion disk extends down to a critical radius
which depends on the accretion rate. 
For rates of the order of the Eddington limit the disk extends down to the
marginally stable orbit, the emission from the disk is intense, 
peaks at energies of $\approx$~1 keV and cools the inner hot cloud (HS).
At low accretion rates, the disk is truncated at large radii and 
the plasma flow becomes geometrically thick (an internal torus 
or quasi-spherical cloud) hot and optically thin in the inner region. 
The disk emission decreases and the peak shifts towards lower energies.
This hot plasma Comptonizes the disk photons and produces the hard 
component of the LS.
Stable hydrodinamical solutions for such a hot flows
have been found for low value of the accretion rate, the so called advection
dominated accretion flow (ADAF) models, and they have been employed to describe
BH state transitions using the mass accretion rate as basic parameter 
(Esin et al. 1998).
The optically thick disk also reflects the hard photons and give rise to 
the reflection component more visible during HS but also present in LS
(Gierlinski et al. 1997, 1999). 
The reflection component shows a hump around 30 keV, 
Fe fluorescence line and K-edge feature around 6-8 keV.
Correlation between reflection and hard component slope, $\nu_B$
and presence of soft component may support 
the view that reflection is due to the disk view by a inner
hot cloud (Gilfanov et al. 1999) and variations are linked to
a change in the internal disk radius.
However there is not full agreement on this correlation,
the VHS does not fit very clearly in this picture as well as
the fact that several XN are seen only in LS, even at
high luminosities.

Alternative models propose that the hot cloud is rather a corona 
envelopping the disk. A uniform corona however is ruled out 
by the relative low level of reflection,
and solutions involving patchy or evaporating coronae have been 
recently proposed (e.g. Malzac et al. 2002).
Another model considers the two basic spectral components arising from
two distinct flows, one the standard Keplerian disk the other a 
sub-Keplerian flow heated 
by shocks within the flow (e.g. at the centrifugal barrier) 
(Chakrabarti $\&$ Titarchuk 1995). 
In any case several important questions on geometry of the flow
(hot inner cloud or corona), hydrodinamics/heating processes (viscous
dissipation or magnetic reconnections) and radiation 
mechanisms (pure thermal or non-thermal processes) remain open
(see reviews by Liang 1998, Zdziarski 2000).

Of particular relevance is the problem of the origin of the hard tail 
during HS or VHS, when the slope of this component is softer than during 
LS and does not show high energy cut off. 
One possibility is that the Bulk Motion Comptonization (BMC)
of disk low-energy
X-rays by free-falling electrons very close to the BH horizon 
is the dominant mechanism of this steep hard tail during HS 
(Laurent $\&$ Titarchuk 1999). BMC model 
predicts correctly the power-law slope but also a high-energy cut off 
($>$ 100-300 keV) which has not been observed yet.
In particular, it has been recently repoted a detailed study of 
all CGRO data concerning \cygx\ in HS from which it appears 
that the source is clearly detected in the energy bands 
1-3 MeV and 3-10 MeV, with fluxes compatible with simple extension 
of the power-law at 10 MeV (Mc Connel et al. 2002, Fig. 3 left).
A competitive model to the BMC is the hybrid thermal/non-thermal model 
(Zdziarski 2000) which ascribes instead this tail to emission from 
an additional non-thermal population of electrons.

Even if the LS hard component seems well explained by
thermal Comptonization plus a reflection component, a contribution from
non-thermal population has also been considered. 
Indeed the recent
compilation of Comptel/CGRO high energy data on \cygx\ by Mc Connel et al. 
(2002) seems to indicate the presence of a hard component at $>$ 600 keV, 
sticking out from the thermal exponential tail, observed from the source in LS.
A similar result was obtained with SIGMA and with Compton/GRO data also
on the bright XN GRO~J0420+32, which was in LS throughout its entire
outburst (Roques et al. 1994).

A recent issue connected with the detections of high energy tails 
is the observed correlation of the LS hard component 
with radio emission interpreted as thick synchrotron 
radiation from the base of jets in the source. 
Recent multi-wavelength observations of BH X-ray binaries, as e.g. 
for XTE~J1550-564,
have shown clear presence of flat-spectrum radio emission
when they are in low state with positive correlation of the
intensities. This suggests a coupling between
radio jets and Comptonization, and possibly contribution
by the accelerated electrons to X-ray emission via inverse 
Compton or even synchrotron processes (Fender 2001).

\begin{figure}
\centering{
\epsfig{figure=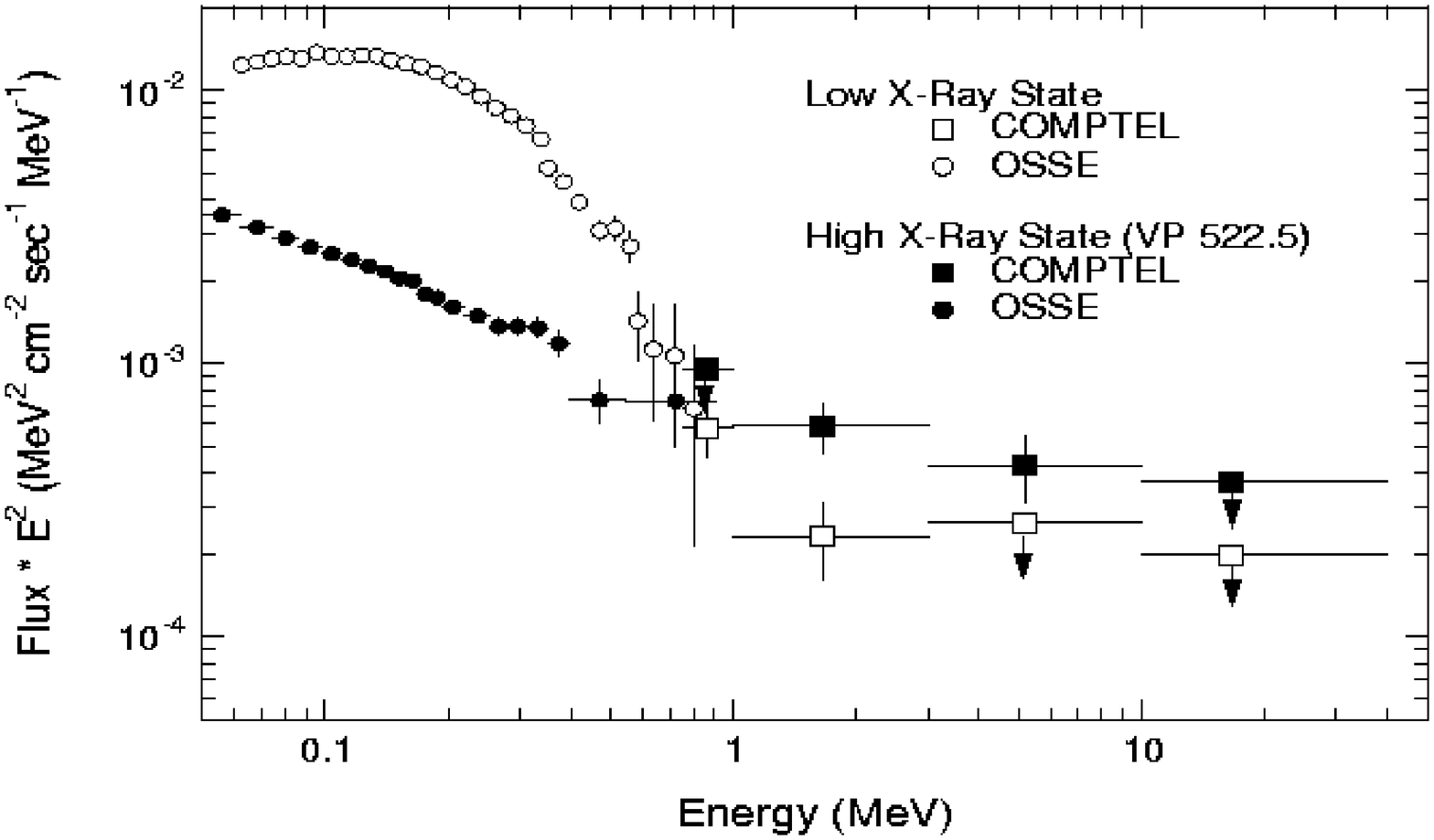, width= 9 cm} ~~~~~~~~
\epsfig{figure=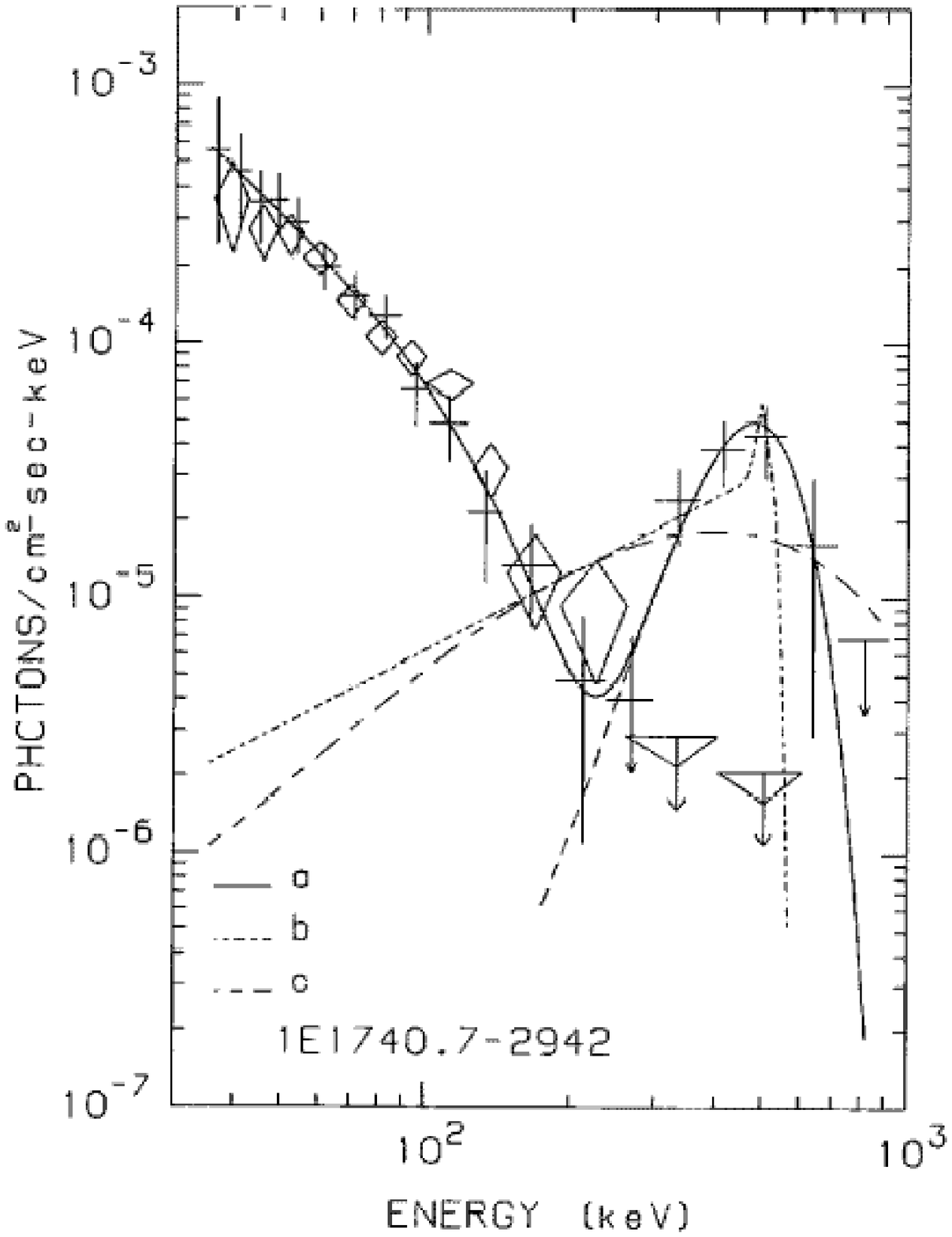, width= 5.5 cm}
}
\caption{High energy Compton-GRO spectrum of \cygx\ 
(from Mc Connel et al. 2002) (left).
Spectrum of 1E~1740.7-2942 during the detection of the high energy feature
(Bouchet et al. 1992) (right).
\label{fig:gbhhen}}
\end{figure}

\section{High Energy Emission Features in GBH}

Apart from the persistent high energy tails 
a number of detections of variable high energy ``features'' 
at $>$ 300 keV, sticking out from the simple extrapolation of 
the continuum emission of GBH have been reported.
They obviously have attracted interest because of their possible link with
electron-positron annihilation, pair plasma and non-thermal radiation 
processes.

The 1~MeV bump of Cyg~X-1 observed with HEAO-3 (Ling et al. 1987)
was the first significant report of such events.
The feature appeared when \cygx\ was in its standard hard state
during a low level of gamma-ray emission (the $\gamma$1 state).
The excess was observed during 14 days in the range 400~keV~-~1.5~MeV.
The feature was never detected again.
On the other hand it was recently reported the detection of a gamma-ray 
transient burst of 10 ks duration, with Ulysses and Konus-Wind 
from a direction compatible with the \cygx\ position (Golenetskii et al.,
IAUC~7840). 

A broad emission feature in the 300-700 keV band was detected with 
the SIGMA/GRANAT telescope in automn 1990 from the hard X-ray 
source and microquasar 1E~1740.7-2942 (Bouchet et al. 1992).
The feature was observed for 1 day at a level of $>$~6~$\sigma$,
could be fitted with a large Gaussian line, and was not compatible 
with a narrow 511~keV line even associated to positronium emission 
(Fig. 3 right). The excess was not present during the observations 
performed 3 days before and the day after the event.
A following detection during fall 1991 at much lower significance level
(4~$\sigma$) (Cordier et al. 1993) was not confirmed by 
simultaneous Compton GRO observations.

Another remarkable result obtained with the soft gamma-ray 
SIGMA/GRANAT telescope was the detection of a high energy variable
line feature from the X-Ray Nova Muscae 1991 (GRS~1124-684), 
a dynamically proven BH, during its
main outburst. The emission line was centered at 480~keV
with flux of 6~10$^{-3}$~ph~cm$^{-2}$~s$^{-1}$ and width of 23 keV,
compatible with instrument resolution (Goldwurm et al. 1992) (Fig.~4).
The variable line was seen few days after the XN reached the
peak of bolometric luminosity when the source was in VHS
for about 20 hours. 
Initially interpreted as red-shifted annihilation line,
many unresolved issues about the possible site of annihilation and 
positron production mechanism remain. 
The positrons could have been formed in the inner hot part of the 
accretion disk and then have annihilated in a colder medium, 
for example in the external part of the disk. However the red-shift 
would imply an annihilation site very close to the BH and
the line should be larger due to Keplerian rotation and to
higher temperatures. Moreover if the positrons are produced 
by $\gamma$-$\gamma$ interactions, with $>$ 511 keV $\gamma$-rays 
from a hot pair plasma or within a jet, an intense $>$ 500 keV 
continuum component should have also been observed.
A number of models were proposed in this frame, 
and recently Kaiser and Hannikainen (2002), making a correlation
with the radio flare which intervened few days after, proposed that 
the line is produced within a jet and the frequency shift is due to 
doppler effect.

A second feature at $\approx$~200 keV seems also present in the spectrum 
and was interpreted as backscattering of 511 keV from the disk edge
(Hua $\&$ Lingenfelter 1993) or as result of distorsion of the 511 keV 
line in a Keplerian disk around a BH (Hameury et al 1994).
A different interpretation was proposed when relevant amount 
of Li was discovered in the secondaries of BH XN. Martin et al (1996) 
proposed that 478 keV line could arise from decay of excited $^7$Li 
produced in $\alpha$-$\alpha$ reactions directly or from Be decay
obtained in spallation processes during the outburst. 
Certainly suggestive, this view encounters difficulties because this 
mechanism would also generate excited $^7$Be which would produce a
line at 430 keV with similar intensity, therefore
combined 478 and 430 lines would not be so narrow for NaI 
scintillators instruments like SIGMA. 
Guessoum $\&$ Kazanas (1999), re-discussing a scenario of Li production
in hot ADAF proposed by Yi and Narayan (1997) and in general in
nucleosynthesis processes in accretion flows around BH, found that 
Li would be depleted by proton collisions or photodissociation in the hot 
flow and would not provide detectable fluxes.
On the other hand they investigated production of 
2.223 MeV gamma-ray line from neutron capture by protons, 
due to the large neutron fluxes obtained in the hot accretion flows
intercepting the secondary.

In any case transient high-energy emission lines of such intensities
remain rare events, 
since neither SIGMA nor CGRO ever detected such features in other 
X-Ray Novae (Goldoni et al. 1999, Grove et al. 1998, Cheng et al. 1998), 
even if these results may be due to the non complete coverage of 
XN in VHS close to peak activity.
Upper limits on high energy lines from GBH reported from CGRO data 
are of the order of 5-15~10$^{-4}$~ph~cm$^{-2}$~s$^{-1}$ for one day
of data integration. 

\section{New Results and Perspectives}

Other recent results obtained in particular with RXTE, Beppo-SAX, Chandra 
and XMM-Newton, show however that the standard picture and 
in particular the assumption that state transitions are driven 
by accretion rate changes, is not fully compatible with the data.

One new important finding of the last years was the discovery with RXTE
of high frequency (60 - 450 Hz) QPO 
in GBH and in particular in the two superluminal 
microquasars GRS~1915+105 and GRO~1655-40 (Strohmayer 2001). 
The involved timescales imply emission site very close to 
the BH horizon and therefore pose strong constraints on the 
radiation mechanisms and possibly on BH spin and mass of the system
(Greiner et al. 2001).

Observations of XN XTE~1550-564 have showns that the source changed state 
on timescale of days and weeks with no clear correlation with accretion rate
(Homan et al. 2001). 
The authors conclude on this basis and other indications that
at least another independent parameter, in addition to mass accretion rate,
drives the state transitions in GBH systems.

Another important recent result is the detection of a peculiar 
low/soft state in the persistent GBH and microquasar GRS~1758-258
(Smith et al. 2001, Miller et al. 2002, Goldoni et al. 2002, Goldwurm et al. 2002) 
and the fact that in this source,
differently than for \cygx, the spectral changes seem correlated to
flux changes rather then being correlated to intensity itself.

All these results open new questions regarding the physics involved in 
GBH which definitely require further studies and deeper
high energy broad band observations.
INTEGRAL, thanks its broad band sensitive range (3 keV - 10 MeV) 
large sensitivity and spectral/imaging capabilitites will 
provide new deep insight in the domain of the astrophysics of
black hole systems. The persistent GBH and the transient GBH are
indeed major target for the mission and we expect that important results
will be obtained from the INTEGRAL observation program of these objects, 
in particular if correlated to multiwavelength observation programs
involving soft X-rays, radio, optical/IR telescopes.

\begin{figure}
\centering{
\epsfig{figure=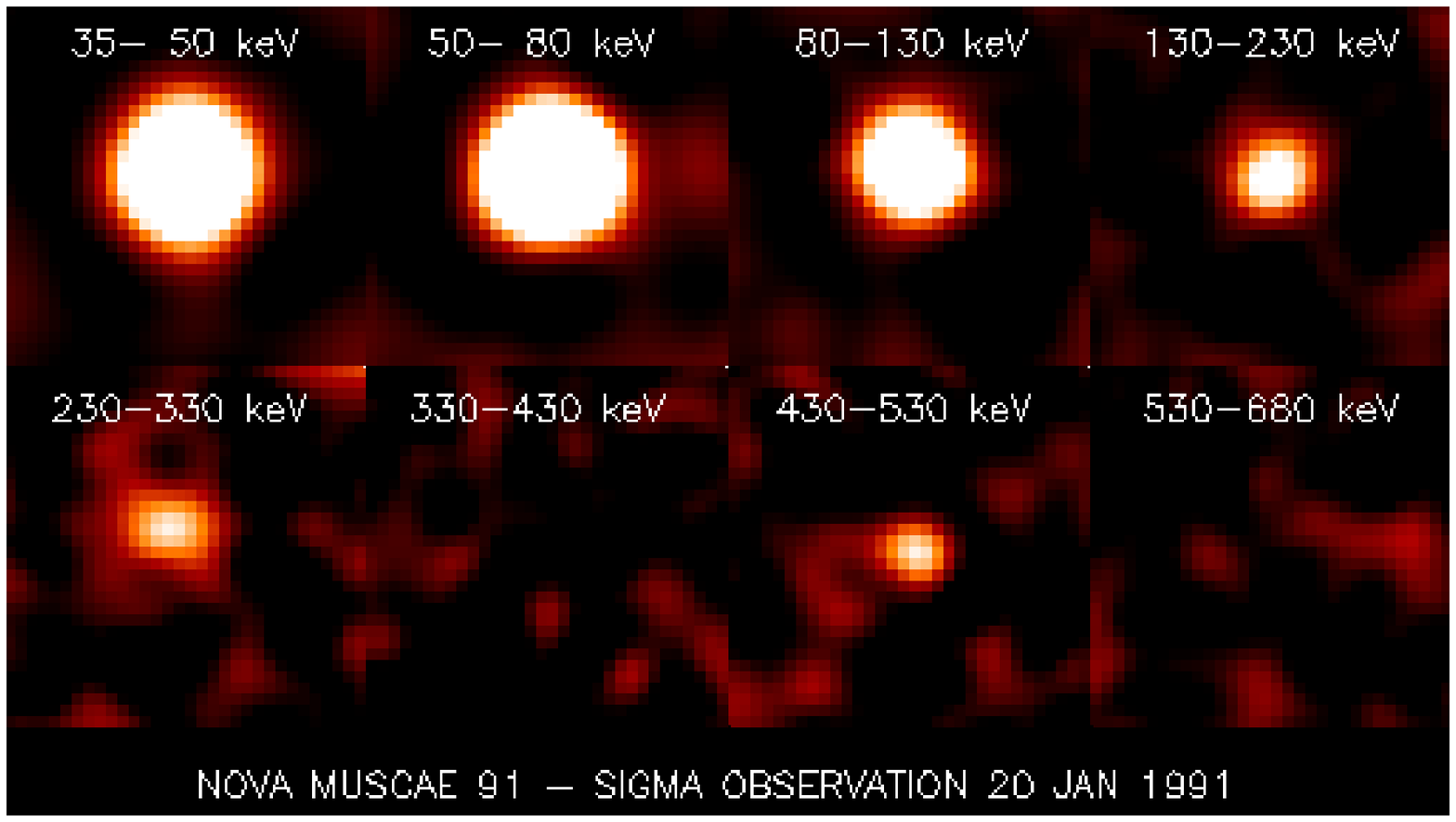, width=8.5 cm} ~~~~~~~~
\epsfig{figure=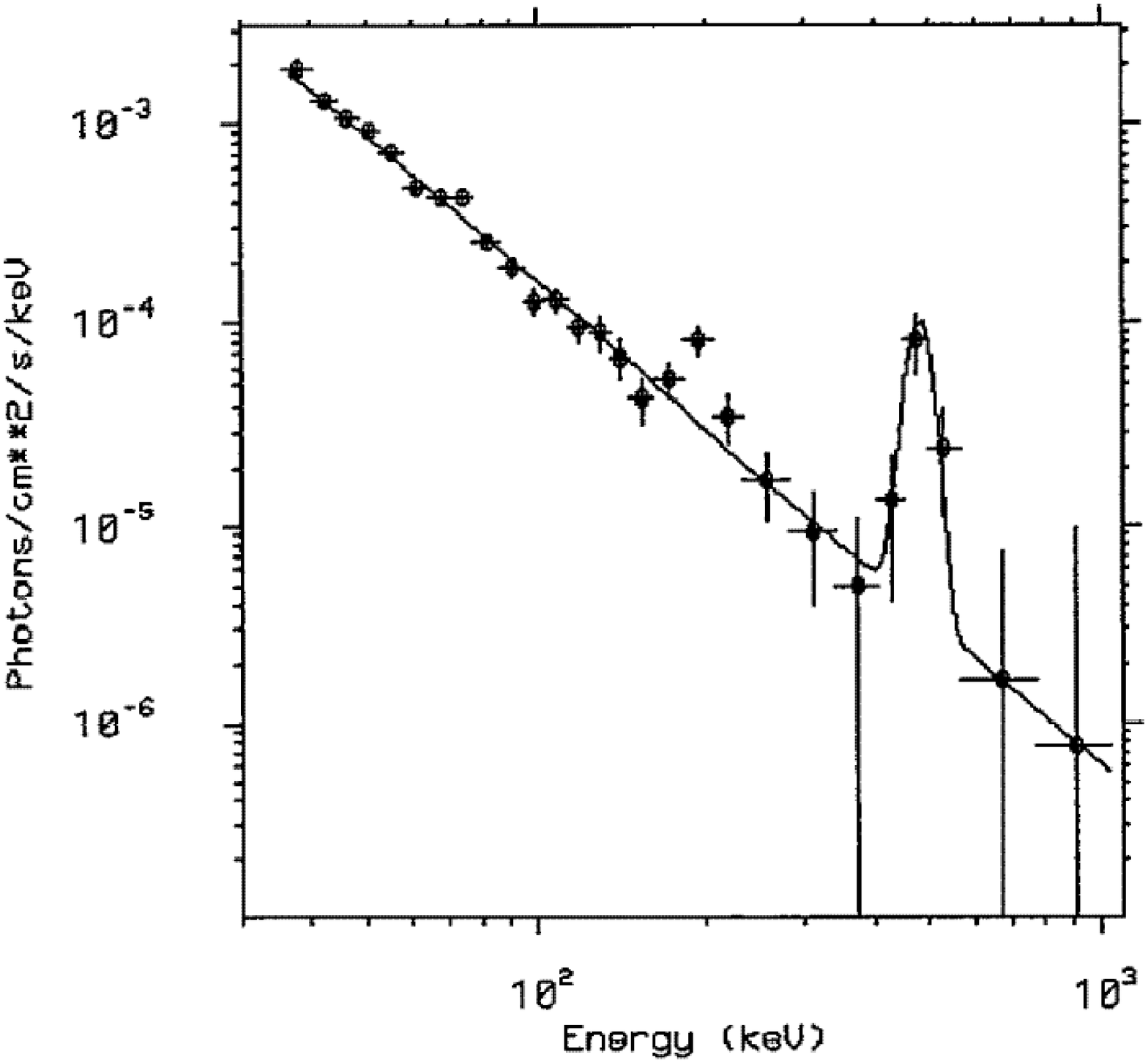, width=5.5 cm}
}
\caption{Gamma-Ray images in different energy bands
of the X-Ray Nova Muscae recorded by the SIGMA/GRANAT 
telescope on 20 Jan 1991 (left). The image in the 430-530 keV 
band shows a significant excess at the source position 
while no signal is found in the lower energy one.
The corresponding photon spectrum of the source
fitted with a power-law and a Gaussian line 
(from Goldwurm et al. 1992) (right).
\label{fig:musca}}
\end{figure}

\end{document}